\documentclass[article, 10pt]{aa}  

\usepackage{amsmath,amssymb,lmodern,mathrsfs,ulem,mathtools,bm}
\usepackage{graphicx}
\usepackage{enumitem}
\usepackage{natbib}
\usepackage[dvipsnames]{xcolor}
\usepackage{makecell}
\usepackage{hyperref}
\usepackage{lipsum}
\usepackage[modulo]{lineno}
\usepackage[caption=false]{subfig}
\usepackage{fancyhdr}
\usepackage{orcidlink}

\hypersetup{
    colorlinks=true,
    citecolor=blue,
    linkcolor=blue,
    filecolor=magenta,      
    urlcolor=cyan,
    pdftitle={Overleaf Example},
    pdfpagemode=FullScreen,
    }

\newcommand{\hide}[1]{}

\newcommand{\grad}{\mbox{\boldmath$\nabla$}}
\newcommand{\drm}{\mathrm{d}}
\newcommand{\mbf}[1]{\boldsymbol{#1}}
\newcommand{\divrm}{\mathrm{div}}
\newcommand{\bhat}{\hat{\mathbf b}}

\definecolor{forest}{RGB}{34,139,34}

\begin{document}


\subtitle{\LARGE Direct calculation of steady-state hydrodynamic solar wind solutions with newtonian viscosity}

\author{
    {Roger B. Scott} \inst{1,4} 
    \orcidlink{0000-0001-8517-4920}
    \and {Stephen J. Bradshaw}\inst{2} 
    \orcidlink{0000-0002-3300-6041}
    \and {Mark G. Linton}\inst{1} 
    \orcidlink{0000-0002-4459-7510}
    \and {Chris Lowder}\inst{3} 
    \orcidlink{0000-0001-8318-8229}
    \and {Leonard Strachan}\inst{1}
    \orcidlink{0000-0002-5425-7122}
    }

\institute{
    US Naval Research Laboratory, Washington, DC, USA \hide{\NRLDSA} \\
    \and Rice University, Houston, TX, USA\\
    \and Southwest Research Institute, Boulder, CO, USA\\
    \and Montana State University, Bozeman, MT, USA}

   \date{}
 
  \abstract
   {{Steady-state solutions to the Navier-Stokes equations are a valuable tool for constructing quasi-steady models of the solar wind and exploring the various factors that affect the fluxes of mass, energy, and momentum into the heliosphere.
   These models typically omit the effects of viscosity, which is assumed to be negligible under most coronal and heliospheric conditions;
   however, the inviscid Navier-Stokes equations are known to admit solutions that are singular at the sonic point, where the solar wind speed becomes equal to the relevant acoustic speed.
   Consequently, inviscid solar wind models require special treatment of the solution near the sonic points, and this has proven to be a significant impediment to efficient modeling of the solar wind.}}
   {In this paper we {revisit the governing hydrodynamic equations for the expanding solar wind, with the inclusion of the viscous stress as defined by the classical (Newtonian) closure, and we show how this inclusion eliminates the singularities that emerge from the inviscid equations.
   This result has been previously reported and used to generate steady-state solar wind profiles from initial conditions in the asymptotic limit (outside of the Sun's gravitational well);
   however, those studies did not include realistic treatments of the inner corona, and generally rejected the prospect of extrapolating solutions outward from the Sun into the heliosphere, which they deemed to be computationally unfeasible.
   Our aim, therefore, is to expand this method to include external heating and optically thin radiative losses and show that solutions can be computed outward from initial conditions near the solar surface, thereby capturing the entire range of scales from below the transition region to the outer heliosphere in a single solution.}} 
   {Our {approach was to cast} the steady-state, field aligned Navier-Stokes equations as a system of five coupled, first order, ordinary differential equations (ODEs) describing the spatial evolution of the mass density, pressure, speed, conductive heat flux, and viscous stress. 
   These equations were then solved {using conventional methods, without any special treatment of the governing equations in the vicinity of the sonic point.
   Physically meaningful solutions were identified by varying the initial conditions at the lower boundary until the solution obtained the correct asymptotic form, which we derived for the particular closures that we employed.}}
   {{The representative solutions that we present here demonstrate the utility and efficiency of this extrapolation method, which is considerably more realistic than commonly used analytical or empirical models.}
   This method provides a direct approach to generating accurate solar wind profiles subject to {observationally} motivated initial conditions near the solar surface, at a fraction of the computational cost of comparable relaxation-based models. 
   The {solutions obtained from} this method can be used to initialize time-dependent simulations, to {generate} large families of steady-state solutions that can then be used to populate the hydrodynamic variables along individual magnetic field lines in {global magnetic field models}, {and to explore how the properties of the quasi-steady solar wind are affected by changes in magnetic geometry and different coronal heating models.}}
   {}


   \maketitle

\section{Introduction}\label{Sec::Introduction}

{
For over half of a century, it has generally been understood that time-steady solar wind streams are subject to critical conditions at the sonic point(s), where the outflowing fluid attains a flow speed equal to the relevant acoustic speed.
This was first discussed by \citet{Parker:1958n}, who predicted the existence of a transonic solar wind from consideration of an inviscid isothermal solar atmosphere, which he argued must attain a nonzero asymptotic radial velocity in order to facilitate an asymptotic reduction of the plasma pressure at large distances from the Sun.
Parker observed that the only solutions whose radial velocity remained finite at a large distance were supersonic solutions, and that of these only one begins as a subsonic outflow at the solar surface.
This transonic solution accelerates monotonically up to the critical point, which represents the only location (for a given temperature) where the flow speed can attain the acoustic speed without generating singular conditions, before increasing to supersonic speed downstream of the critical point, as needed to enforce the requirement of vanishing pressure in the extended heliosphere.
}

{
The existence of such critical points, and the constraints that they imply, are persistent features of all inviscid hydrodynamic solar wind models.
For example, the generalization of Parker's isothermal solar wind to a larger class of analytical polytropic solutions \citep{KoppHolzer:1976} exhibits the same fundamental structure despite the inclusion of adiabatic cooling in the polytropic energy equation.
Similarly, numerical solutions to the inviscid nonideal fluid equations -- including such effects as thermal conduction and external heating and cooling, which often preclude analytical solutions -- all exhibit critical behavior at the sonic point.
}

{
One of the more direct approaches to navigating these critical conditions was described by \citet{Noble:1963}, who solved the inviscid hydrodynamic equations with thermal conduction by numerical integration from initial conditions at $1\,\rm AU$.
Their method of solution was to treat the conditions at the sonic point as an interior boundary condition and to vary the initial heat flux until the critical condition was satisfied at that boundary.
They then expanded the governing equations near the critical point using L'Hopital's rule in order to minimize numerical errors, thereby allowing their integration to proceed through the critical point, before reverting to the nonlinear fluid equations in order to continue the integration into the subsonic lower corona.
}

{
Other, more novel approaches have also been developed for navigating or circumventing the singular conditions at the critical point.
For example, \citet{Melia:1988} performed a change of variables that removed the singularity from the momentum equation, but which then necessitated that the new primitive variable obtain a particular value at the location of the critical point in order for the solution to be continuous there -- hence, the singular conditions at the critical point were removed but the need to explore the space of initial conditions in order to satisfy the requisite physical constraints at the sonic point persisted.
Conversely, \citet{Hammer:1982} opted to calculate solutions from initial conditions at the base of the corona, again using the singular conditions at the critical point as a boundary condition, but then relied on previous results from \citet{Roberts:1972} to map the plasma properties at the critical point to conditions in the asymptotic region, thereby further constraining the initial conditions at the solar surface without actually performing any calculations in the region downstream of the critical point.
}

{
While these techniques have allowed a number of authors to calculate self-consistent solar wind profiles with nonideal effects, the necessity of selecting between different fluid models in different spatial domains and the need to simultaneously satisfy the constraints at the sonic point and any other relevant boundary conditions imply considerable additional complexity beyond the core task of integrating the governing equations.
However, as demonstrated by \citet{Scarf:1965} and \citet{WLC:1966}, when the effects of viscosity are included in the governing equations, the sonic point ceases to be a location of particular interest, as the viscous hydrodynamic equations governing the solar wind contain no singularities.
Hence, while the viscous Navier-Stokes equations are in some respects more complex than their inviscid counterparts, their solutions can be computed using comparatively simple computational frameworks.
}

{
Despite the absence of critical points in the viscous Navier-Stokes equations, it is still possible to compute solutions that are mathematically valid but which do not correspond to physically meaningful descriptions of the solar wind \citep{Parker:1965, Dahlberg:1970}.
In particular, it has been shown that the viscous equations can lead to spurious solutions that either (a) exhibit unphysical behavior (i.e., zero pressure at finite distance from the sun), or (b) reduce to breeze solutions (i.e., zero velocity at large distance from the sun).
Hence, when computing these viscous solutions it is necessary to impose additional constraints \citep{Summers:1980} in order to ensure that the solutions exhibit reasonable fluxes of energy and momentum and obtain appropriate asymptotic forms at large distance from the sun (outside of the Sun's gravitational well).
Nonetheless, the existence of viscous transonic solar wind solutions and the conceptual framework for extracting them from the larger solution space has been well established since the early 1980s.
}

{
Unfortunately, even with this framework, studies of the viscous solar wind face an additional computational challenge in that the viscous Navier-Stokes equations exhibit the properties of a {stiff} system, particularly when integrated in the direction of increasing speed.
For this reason, solutions have typically been computed by choosing initial conditions that are consistent with the asymptotic constraints at some arbitrary location in the asymptotic region, and then integrating in the direction of decreasing distance from the sun.
This approach has the benefit that the asymptotic boundary conditions are enforced by construction; however, it suffers from the inherent problem that the values of the basic plasma primitives are not known in the asymptotic region, and must be systematically varied until a transonic solution is found that yields reasonable plasma conditions in the lower corona.
Hence, just as with the inviscid models, satisfactory solutions can only be found by exploring the space of initial conditions. 
}

{
This approach to solving boundary value problems (BVPs) using the method of shooting -- recasting the BVP as an initial value problem (IVP) and then exploring the space of initial conditions to satisfy the original BVP -- can be computationally intensive, and would have been particularly onerous during the 1970s and 1980s when interest in viscous solar wind solutions was at its highest.
The limited computing resources of that time likely contributed to a decline in research in that area, which has seen little development since the mid 1980s.
Consequently, and in the absence of readily available tools for efficiently computing self-consistent, time-steady solar wind solutions with realistic thermodynamics, the majority of modern solar wind models continue to rely on simple isothermal and polytrpic fluid models \citep[e.g.,][]{Dakeyo:2022, Lowder:2024a}, empirical models \citep[e.g., the famous WSA model,][]{Wang:1990j}, or computational intensive time-dependent and pseudo-time-dependent models \citep[e.g.,][]{Cranmer:2007, Grappin:2011, Wang:2012a}.
At present, there are no readily available solar wind models that can rapidly generate solutions to the governing equations of a time-steady solar wind for given specifications of the external heating rates and magnetic geometry, with realistic physics to cover the range of scales that are necessary to accurately describe the solar wind from its origin near the solar surface to its terminus at the heliopause.
}

{
In this paper we present a remedy to this absence of a rapid and robust tool for calculating steady-state solar wind solutions by revisiting the viscous Navier-Stokes equations, which can now be solved with relative ease using modern computational methods and resources.
We begin in Section \ref{Section::ODE} by re-deriving the governing equations for a conducting, viscous, time-independent solar wind with external heating and radiative cooling, and we explore the formalism behind the presence and absence of critical points in the various limits.
In Section \ref{Section::Results} we then describe solutions to the systems of ODEs in two familiar limits: the isothermal limit (for comparison to the original Parker solution) and the general case with parameterized external heating and cooling.
In both cases we discuss the existence (or absence) of a physically meaningful asymptotic form (which we discuss further in Appendix \ref{App::ASols}) and the corresponding boundary conditions that we impose.
We then go on to discuss implications of this method in Section \ref{Section::Discussion}, before summarizing our results in Section \ref{Section::Conclusion}.
}

\section{Methods}\label{Section::ODE}

\subsection{Governing equations}

{
For the purpose of this demonstration we adopt a single-temperature\footnote{The generalization to independent ion and electron temperatures does not materially affect this analysis.} fluid model of the solar wind}, with phenomenological heating and forcing, optically thin radiation, conductive heating, viscous heating, and viscous forcing. 
{We assume that the plasma is in a steady state, which is described by the following system of time-independent, coupled differential equations:}
\begin{gather}
    \divrm (\rho u) = 0, \label{Eqn::ConsContinuity}\\
    \divrm (\rho u^2) = \dot{\mathcal{P}} - \partial_s P - \divrm (\sigma), \label{Eqn::ConsMomentum}\\
    \divrm (P u) = (\gamma - 1) \left(\dot{\mathcal{Q}} - P\, \divrm (u) - \divrm (f) - \sigma \partial_s u \right) \label{Eqn::ConsEnergy}.
\end{gather}
Here $\divrm (\circ) \equiv (1/A) \partial_s (A\, \circ)$ is the field-aligned divergence operator (acting on $\circ$) and $\partial_s$ is the directional derivative along a magnetic flux tube with cross-sectional area $A(s)$.
The density of the fluid is $\rho$, while $u$ is its field-aligned velocity (speed), and $P$ is the total pressure (sum of contributions from ions and electrons).
The ratio of specific heats is $\gamma$, which we take to be a parameter of the model, {with $\gamma\rightarrow1$ corresponding to an isothermal condition while $\gamma\rightarrow5/3$ corresponds to systems of monatomic gasses with adiabatic heating and cooling.}

The terms on the right-hand side (RHS) of the momentum equation \eqref{Eqn::ConsMomentum} are the external forcing (rate of momentum deposition), pressure gradient, and divergence of the viscous stress, while the terms on the RHS of the energy equation \eqref{Eqn::ConsEnergy} are the external heating and cooling rates (rate of energy deposition), and the internal heating and cooling rates due to adiabatic expansion, thermal conduction, and viscosity.
The external forcing ($\dot{\mathcal{P}}$) comprises both the gravitational force and any additional phenomenological momentum deposition, while the external heating and cooling ($\dot{\mathcal{Q}}$) accounts for both phenomenological heating and optically thin radiative losses.
The details of the external heating and cooling are given in section \ref{Section::Results}.

The viscous stress and conductive heat flux are taken, {respectively}, to be
\begin{gather}
    f = - \kappa(T) \partial_s T \label{Eqn::HfxClsr}\\ 
    \sigma = - \mu(T) \partial_s u \label{Eqn::VscClsr},
\end{gather}
where $\mu(T)$ is the coefficient of dynamic viscosity and $\kappa(T)$ is the thermal conductivity.
The temperature $T$ is {related to the} total pressure by the ideal gas law as
\begin{equation}\label{Eqn::IGas}
P = 2 n k_B T,
\end{equation}
{and we note that the temperature gradient can be} expanded in terms of the primitive variables as
\begin{equation}
    \partial_s T = (T/P)\, \partial_s P - (T/\rho)\, \partial_s \rho.
\end{equation}
Here $n = \rho / m_i$ is the number density of ions (or electrons), $m_i$ is the ion mass, and $k_B$ is Boltzmann's gas constant.

{The closure of $\sigma$ through equation \eqref{Eqn::VscClsr} and its contribution to the energy and momentum equations, \eqref{Eqn::ConsMomentum} and \eqref{Eqn::ConsEnergy}, differ somewhat from the field-aligned projections of the viscous, anisotropic Navier-Stokes equations as discussed by \citet{Summers:1980} and others.\footnote{
A more exact construction is derived in Appendix \ref{App::Vsc} from consideration of the full anisotropic viscous stress tensor.}
The expressions used here are a convenient simplification, whose form resembles the Newtonian closure of the isotropic viscous stress tensor in 3D, and which is more approachable for the purpose of constructing asymptotic solutions.
Moreover, the classical definitions of the viscous stress and conductive heat flux \citep{Braginskii:1965, Holzer:1986a, Fitzpatrick:2015} are known to permit energy fluxes that  exceed the physical limits implied by the free-streaming of electrons and structural constraints on the viscous stress tensor \citep[see, e.g.,][]{Bale:2013, Landi:2014}.
This can be prevented through the application of flux-limiters (see Appendix \ref{App::FLims} for details); 
however, for the sake of simplicity we have elected not to employ those regularizations in the current model.
In any event, any such closure -- either with or without flux-limiting regularization -- is, at best, an approximation of the higher moments of the velocity distribution function in the collisionless regime, and it is not clear what benefit the exact anisotropic form holds over the simplifications used here.}

\subsection{Matrix representation}

Eqs. \eqref{Eqn::ConsContinuity} - \eqref{Eqn::VscClsr} constitute a system of five coupled ODEs in the independent variable $s$, and we wish to cast them in such a way that the state vector $\mbf{U}(s)$ can be expressed in terms of the integral
\begin{equation}\label{Eqn::ODE_IVP}
    \mathbf{U}(s) = \mathbf{U}(s_0) + \int_{s_0}^{s} \partial_s\mbf{U}(\mbf{U}(s')) \,\drm s',
\end{equation}
where $\partial_s \mbf{U}(\mbf{U}(s))$ is the derivative of $\mbf{U}(s)$ with respect to $s$, expressed as a pure function of $\mbf{U}(s)$.
If such a function exists and is well behaved, $\mbf{U}(s)$ can be determined from $\mbf{U}(s_0)$ as an IVP. 

If we assume that $\dot{\mathcal{Q}}$ and $\dot{\mathcal{P}}$ do not carry hidden dependence on $\partial_s \mathbf{U}$ (i.e., they can be expressed entirely as functions of the primitive variables or external parameters), the governing system of ODEs can be written as
\begin{equation}
    \uuline{M} \cdot \partial_s \mathbf{U} = \mathbf{K},
\end{equation}
where $\uuline{M}$ is a matrix of coefficients, $\mbf{K}$ is a vector of inhomogeneous source terms, and both are understood to be independent of the elements of $\partial_s \mbf{U}$ \citep[see][and references therein]{Tarr:2024}.
Then, so long as $\uuline{M}$ has a well-posed inverse, the derivative of the state vector is given by
\begin{equation}
    \partial_s \mbf{U}(s) = \uuline{M}^{-1} \cdot \mbf{K}.
\end{equation}
{Notwithstanding the various novel methods discussed in section \ref{Sec::Introduction},} the integrability of Eq. \eqref{Eqn::ODE_IVP} as an IVP depends, therefore, on whether the coefficient matrix is invertible.

To proceed, we first expand Eqs. \eqref{Eqn::ConsContinuity}--\eqref{Eqn::VscClsr} and collect terms in the elements of $\partial_s \mbf{U}$, and then normalize each equation so that the source terms have units of inverse length.
Taking our state vector to be 
\begin{equation}
    \mbf{U} = \left(\rho, P, u, f, \sigma\right)
\end{equation}
the source vector can be written as
\begin{equation}
    \mathbf{K} = 
   \begin{pmatrix}
    -\partial_s \ln A \\
    \frac{\gamma - 1}{Pu}( \dot{\mathcal{Q}} - f\, \partial_s \ln A ) - \gamma\, \partial_s \ln A \\
    \frac{1}{\rho u^2} \left ( \dot{\mathcal{P}} - \sigma \partial_s \ln A \right ) \\
    {f/(\kappa T)} \\
    {\sigma/(\mu u)}
\end{pmatrix}. \label{Eqn::GenSrcVec}
\end{equation}
It then follows that the coefficient matrix is given by
\begin{equation}\label{Eqn::GenCoeff}
    \uuline{M} = 
    \begin{pmatrix}
        & \frac{1}{\rho} & 0 & \frac{1}{u} & 0 & 0 & \vspace{0.2em}\\
        & 0 & \frac{1}{P} & \frac{\gamma P + (\gamma - 1) \sigma}{P u} & \frac{\gamma - 1}{Pu} & 0 & \vspace{0.2em}\\
        & 0 & \frac{1}{\rho u^2} & \frac{1}{u} & 0 & \frac{1}{\rho u^2} & \vspace{0.2em}\\
        & \frac{1}{\rho} & \frac{-1}{P} & 0 & 0 & 0 & \vspace{0.2em}\\
        & 0 & 0 & \frac{-1}{u} & 0 & 0 &
    \end{pmatrix}.
\end{equation}

In order to make the coefficient matrix more tractable, it is convenient to now introduce the diagonal normalization matrix $\uuline{N} = diag \left(\rho, P, u, P u, \rho u^2 \right )$, which can be inserted into our matrix equation so that 
\begin{equation}
    \partial_s \mbf{U} = \uuline{N} \cdot \left ( \uuline{M} \cdot \uuline{N} \right )^{-1} \cdot \mbf{K}.
\end{equation}
Since the entries along the diagonal of $\uuline{N}$ have the same units as the entries in each column of $\uuline{M}$, the elements of $\uuline{M}\cdot\uuline{N}$ are dimensionless, and since $\uuline{M}$ will be invertible if (and only if) $\uuline{M}\cdot\uuline{N}$ is invertible, it suffices to determine whether the determinant of $\uuline{M}\cdot\uuline{N}$ exists in order to ascertain whether Eq. \eqref{Eqn::ODE_IVP} is integrable.
Moreover, by eliminating specific rows and columns of $\uuline{M}\cdot\uuline{N}$ we can infer the effects of removing individual primitive variables from the underlying system of governing equations, meaning that Eq. \eqref{Eqn::GenCoeff} also summarizes the properties of several common simplifications relating to viscosity and thermal conduction.

In the following sections we explore a few relevant limits, including (\ref{sub::inv_ply}) an inviscid, polytropic plasma; (\ref{sub::inv_cnd}) an inviscid plasma with nonideal heating; and (\ref{sub::vsc_iso}) a viscous, isothermal plasma, in order to build intuition for the problem. 
We then return in Section \ref{sub::vsc_cnd} to the full system of equations for a viscous plasma with nonideal heating and cooling, which is our primary interest here.

\subsection{Inviscid polytropic limit}\label{sub::inv_ply}

We begin with the inviscid limit, with a polytropic equation of state, in which the energy equation takes the form $\partial_s \ln P = \gamma \partial_s \ln \rho$.
This is achieved by setting $\dot{\mathcal{Q}} = f = \sigma = 0$.
Since $f$ and $\sigma$ no longer depend on the other primitives in this approximation, we remove the 4th and 5th rowsand columns from $\uuline{M}$, $\uuline{N}$, $\mbf{U}$, and $\mbf{K}$.
The scaled and reduced coefficient matrix is then 
\hide{gradient of the reduced state vector then becomes
\begin{equation}
    \uuline{N}^{-1} \cdot \partial_s \mathbf{U} \rightarrow \left ( \frac{1}{\rho} \partial_s \rho, \frac{1}{P} \partial_s P, \frac{1}{u} \partial_s u \right ),
\end{equation}
and the reduced source vector becomes
\begin{equation}
    \mathbf{K} \rightarrow \left (
    -\partial_s \ln A, -\gamma \partial_s \ln A, \frac{1}{\rho u^2} \dot{\mathcal{P}} \right ).
\end{equation}
The scaled coefficient matrix in this case is then}
\begin{equation}
    \uuline{M} \cdot \uuline{N} =
    \begin{pmatrix}
        & 1  & 0                  &  1      & \\
        & 0  & 1                  &  \gamma & \\
        & 0  & \frac{P}{\rho u^2} &  1      &
    \end{pmatrix}
\end{equation}
and its determinant is
\begin{equation}\label{Eqn::det_inv_ply}
    {\rm det} \left ( \uuline{M} \cdot \uuline{N} \right ) = \frac{\rho u^2 - \gamma P}{\rho u^2}.
\end{equation}

This construction demonstrates the {main difficulty encountered when attempting to integrate the inviscid Navier-Stokes equations through the sonic point: namely, that $\partial_s \mbf{U}$ {becomes} unbounded (or, at least, poorly defined) as $\rho u^2 \rightarrow \gamma P$.
This issue can be resolved by constructing the full expression for $\partial_s u$ and requiring that the numerator of that expression vanish at least as rapidly as the denominator, at which point one can apply L'Hopital's rule to construct a linearized version of the governing equations in the viscinity of the sonic point; 
however, the computational architecture required to enforce such constraints adds additional complexity that we prefer to avoid.
Note that the same consideration} also applies in the inviscid isothermal limit, which can be constructed from the inviscid polytropic limit by setting $\gamma \rightarrow 1$.

\subsection{Inviscid limit with nonideal heating}\label{sub::inv_cnd}

We can relax the polytropic condition and (re)introduce thermal conduction and external heating and cooling in the inviscid limit by setting $\sigma=0$ but leaving all other terms intact.
In this case, the scaled and reduced coefficient matrix is 
\hide{gradient of the reduced state vector is
\begin{equation}
    \uuline{N}^{-1} \cdot \partial_s \mathbf{U} \rightarrow \left ( \frac{1}{\rho} \partial_s \rho, \frac{1}{P} \partial_s P, \frac{1}{u} \partial_s u , \frac{1}{P u} \partial_s f\right ),
\end{equation}
while the inhomogeneous source vector is
\begin{equation}
    \mathbf{K} = 
    \begin{pmatrix}
    -\partial_s \ln A \\
    \frac{(\gamma - 1)}{P u}( \dot{\mathcal{Q}} - f \partial_s \ln A ) - \gamma \partial_s \ln A \\
    \frac{1}{\rho u^2} \dot{\mathcal{P}} \\
    {f}/{(\kappa T)}
\end{pmatrix}.
\end{equation}
The scaled coefficient matrix is then}
\begin{equation}
    \left ( \uuline{M} \cdot \uuline{N} \right ) = 
    \begin{pmatrix}
        & 1  & 0                  &  1       & 0            & \\
        & 0  & 1                  &  \gamma  & (\gamma - 1) & \\
        & 0  & \frac{P}{\rho u^2} &  1       & 0            & \\
        & 1  & -1                 &  0       & 0            &
    \end{pmatrix},
\end{equation}
and its determinant is
\begin{equation}
    {\rm det} \left ( \uuline{M} \cdot \uuline{N} \right ) = (\gamma - 1) \frac{\rho u^2 - P}{\rho u^2}.
\end{equation}

As before, the existence of an inverse is predicated on the determinant of $\uuline{M}\cdot\uuline{N}$, which, as in the inviscid isothermal case, vanishes as $\rho u^2 \rightarrow P$. 
Additionally, the presence of a heat flux causes the inverse to be ill-defined when $\gamma \rightarrow 1$, since that limit implies a constant temperature and, therefore, zero conductive heat flux, meaning that $\partial_s f$ cannot be determined from the energy equation in the case of $\gamma=1$.
{As it happens, this singular condition cannot be circumvented by any clever treatment of the governing equations near the singular point, since the determinant does not vanish at a point, but is everywhere zero. 
The isothermal limit cannot, therefore, be constructed as a limiting case of the conducting limit except by explicitly removing all terms associated with the conductive heat flux.}

\subsection{Viscous isothermal limit}\label{sub::vsc_iso}

While it is not possible to construct a viscous polytropic limit (the viscous heating precludes a polytropic energy equation) we can construct a viscous isothermal system by {leaving $\sigma$ intact while} setting $\gamma \rightarrow 1$ and removing the rows and columns associated with $f$.
{In this limit,} the scaled and reduced coefficient matrix is
\begin{equation}
    \left ( \uuline{M} \cdot \uuline{N} \right ) =
    \begin{pmatrix}
        & 1  & 0                       &  1      & 0 & \\
        & 0  & 1                       &  1      & 0 & \\
        & 0  & \frac{P}{\rho u^2}      &  1      & 1 & \\
        & 0  & 0                       & -1      & 0 &
    \end{pmatrix},
\end{equation}
and its determinant is 
\begin{equation}
   {\rm det} \left ( \uuline{M} \cdot \uuline{N} \right ) = 1.
\end{equation}

{Apparently, the inclusion of the viscous stress in the isothermal fluid model dictates that the inverse of the coefficient matrix remain finite for all possible state vectors, meaning that a solution can always be found from direct integration of the governing system of ODEs.
This is a direct consequence of the Newtonian closure for the viscous stress, per Eq. \eqref{Eqn::VscClsr}, from which it follows that the velocity gradient is a simple function of $\sigma$ and $T$, which is independent of $u$ and which is always well posed for finite $\sigma$ and positive $T$.}

\subsection{viscous limit with nonideal heating}\label{sub::vsc_cnd}

Finally, we return to the full set of governing equations as given in Section \ref{Section::ODE}, {including all of the contributions from the viscous stress and conductive heat flux.} 
With all of the terms in Eq. \eqref{Eqn::GenCoeff} included, the scaled coefficient matrix is
\begin{equation}\label{Eqn::GenCoeffNrm}
    \uuline{M} \cdot \uuline{N} = 
    \begin{pmatrix}
       & 1  & 0                  &  1      & 0            &  0 & \\
       & 0  & 1                  &  \frac{\gamma P + (\gamma-1)\sigma}{P} & (\gamma - 1) &  0 & \\
       & 0  & \frac{P}{\rho u^2} &  1      & 0            &  1 & \\
       & 1  & -1                 &  0      & 0            &  0 & \\
       & 0  & 0                  & -1      & 0            &  0 &
    \end{pmatrix},
\end{equation}
and its determinant is
\begin{equation}
    {\rm det} \left ( \uuline{M} \cdot \uuline{N} \right ) = \gamma-1.
\end{equation}
{As with the viscous isothermal case, the inclusion of the viscous stress necessarily suppresses any speed-dependent singular conditions, since $\partial_s u$ is always well defined for reasonable values of $\sigma$ and $\mu(T)$ (assuming $\gamma>1$).

It may, at first, seem surprising that these equations contain no singular conditions.
After all, the simple relationship that governs $\partial_s u$ says nothing about the stability of the other primitive variables.
It is instructive, therefore, to inspect the structure of the scaled coefficient matrix, as given in Eqn. \eqref{Eqn::GenCoeffNrm}.
Since the bottom-most row and right-most column each contain only a single entry, the determinant of this matrix reduces to a single term, which is proportional to the determinant of a lesser $3\times3$ sub-matrix, whose elements are taken from the first, second, and fourth rows and columns of $\uuline{M}\cdot\uuline{N}$.
Since the only terms that depend on the primitive variables are in the third row and column of $\uuline{M} \cdot \uuline{N}$, they do not contribute to the calculation.\footnote{
This structure generalizes to any variant on the Newtonian closure, including the field-aligned projections of the anisotropic viscous stress discussed in Appendix \ref{App::Vsc}.}
}


\section{Results}\label{Section::Results}

\subsection{Model parameters}

In the previous section we described the conditions under which the field-aligned Navier-Stokes equations can be integrated as an initial value problem without {consideration of singular conditions,} a consequence of the introduction of the viscous stress.
Here we present the results of this integration for two useful examples, the viscous isothermal limit, which is a direct analog to Parker's isothermal wind, and the viscous solution with nonideal heating and cooling, which contains the main physical processes necessary to describe the {observed properties of the quasi-steady solar wind.
Solutions to both sets of equations were computed from initial conditions at the solar surface using numerical routines that we developed for this purpose. 
These routines rely on conventional numerical methods, and are available as part of the HelioSHOT project \citep{HelioSHOT_zenodo}, which we discuss in more detail in Appendix \ref{App::num}.}

In both cases we assume spherically symmetric expansion, {so that $s$ is equivalent to the heliocentric radius ($r$) and we choose $s=0$ to correspond to the center of the sun, with the solar surface placed at $s=R_\odot=7\times10^{10}\,\rm cm$.
The area is defined to increase with the square of the distance from the sun, $A(s) \sim s^2$, consistent with a radial magnetic field. 
The external forcing is taken to be the gravitational force, with no additional phenomenological momentum deposition, hence 
\begin{equation}
    \dot{\mathcal{P}} = \rho g_\odot R_\odot^2 / s^2,
\end{equation}
where $g_\odot = -27.4\times10^{3}\,\rm cm\, s^{-2}$
is the gravitational acceleration at the solar surface.
}

{
For the viscous conducting model we set the external heating and cooling to be 
\begin{equation}
    \dot{\mathcal{Q}} = \mathcal{H}(s) + \mathcal{R}(n, T),
\end{equation}
where 
\begin{equation}
    \mathcal{H}(s) = \mathcal{H}_0 \exp(1-s/R_\odot) / A(s)
\end{equation}
is an exponentially decaying phenomenological heating function with amplitude 
$\mathcal{H}_0 = 2\times10^{-6}\,\rm erg\, cm^{-3}\, s^{-1}$, while 
\begin{equation}
    \mathcal{R}(n, T) = -n^2 \Lambda(T)
\end{equation}
is the rate of radiative energy loss.
The emissivity, $\Lambda(T)$, is modeled as a piece-wise continuous power-law function, as described by \citet{Klimchuk:2008}, which represents optically thin emissivity from the solar corona and upper chromosphere.
}

{
The coefficients of dynamic viscosity and thermal conduction follow the usual Spitzer-H\"arm formulation \citep{Spitzer:1953, Braginskii:1965}, and are given here as 
\begin{equation}
    \kappa(T) = \kappa_6 T_6^{5/2}
\end{equation}
and
\begin{equation}
    \mu(T) = \mu_6 T_6^{5/2},
\end{equation}
where $T_6 = T / 10^6\,\rm K$ is the plasma temperature as a fraction of $1\, \rm MK$.
The values 
$\kappa_6 = 8.12\times10^8\, \rm erg \, cm^{-1} \, s^{-1} \, K^{-1}$ and 
$\mu_6 = 1.68\times10^{-1} \, \rm g \, cm^{-1} \, s^{-1}$
represent the combined conductivity of both ions and electrons and the ion viscosity along the direction of the magnetic field, both measured at $T=1\,\rm MK$.
Note that the usual factor of 4/3 has been folded into $\mu_0$, along with a factor of 1/20 from the (assumed constant) Coulomb Logarithm.}

\subsection{Viscous isothermal solutions}\label{sub::num_iso}

While the isothermal limit can be achieved by setting $f=0$ and $\gamma=1$ and then advancing the governing equations for $\rho, P, u$, and $\sigma$, it is more expedient to define the pressure directly in terms of the density as $P = \rho C_s^2$, where $C_s^2 = 2 k_B T / m_i$ is the square of the (constant) isothermal sound speed, which we set to $C_s = 200\,\rm km\, s^{-1}$ for this example.
We are then left with $\rho, u$, and $\sigma$ as the only primitive variables.\footnote{It is also possible to replace the differential form of the continuity equation with a conservation law; however, we find it convenient to retain this equation since the numerical result can be used to characterize the accuracy of the solution.}

For pedagogical purposes we first define the inviscid problem, including the ODE formulation and Parker's analytical solution, which can be compared directly to the viscous variant.
In the inviscid case, the $\rho$ and $u$ evolve according to
\begin{gather}
    \partial_s u = \frac{u}{u^2 - C_s^2}\left ( C_s^2 \partial_s \ln A  + \dot{\mathcal{P}}/\rho \right), \label{Eqn::dsu_invscd}\\
    \partial_s \rho = - \rho \left ( \partial_s \ln u + \partial_s \ln A \right ),
\end{gather}
where we have left the expression for $\partial_s \rho$ in its implicit form (omitting the substitution of $\partial_s u$ from the preceding equation).
{As previously discussed,} Eq. \eqref{Eqn::dsu_invscd} {becomes singular} as $u^2 \rightarrow C_s^2$, {making it} difficult to solve by direct integration;
however, {when $\dot{\mathcal{P}}$ is conservative and} can be written in terms of a potential {(in this case, the gravitational potential)}, Eq. \eqref{Eqn::dsu_invscd} can be anti-differentiated to give
\begin{equation}
    \partial_s u^2 - C_s^2 \partial_s \ln u^2 = C_s^2 \partial_s \ln A^2 - 2 \partial_s \Phi, 
\end{equation}
where $\dot{\mathcal{P}} = - \rho \partial_s \Phi$. 
This results in an equation for the Mach number $(M = u / C_s)$, which is a perfect differential that can be integrated analytically to give
\begin{equation}
    M^2 - \ln M^2 + 2 \Phi / C_s^2 - \ln A^2 = \mathcal{G},\label{Eqn::Parker}
\end{equation}
where $\mathcal{G}$ is a constant of integration.
This is the well-known Parker solution, which is a particular example of de Laval's equation. 

\begin{figure}[ht]
    \centering
    \includegraphics[width=3.5in]{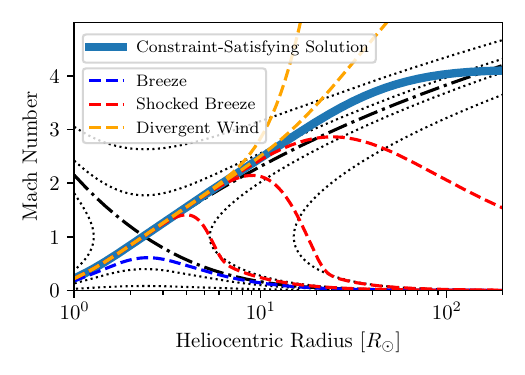}
    \caption{Comparison of the viscous isothermal solutions to the analytical (inviscid) Parker solution. {The black curves represent contours of De Leval's equation \eqref{Eqn::Parker}. Of these, only the black dashed-dotted curve(s) exhibits a sonic point at the allowed critical point, consistent with Parker's description of a transonic solar wind. The variously colored blue, yellow, and red curves are numerical solutions to the viscous isothermal equations. The dashed blue curve is a subsonic breeze solution for which to $u\rightarrow0$ as $s\rightarrow\infty$. The dashed yellow (divergent wind) and red (shocked breeze) curves are transonic solutions that exhibit either $u\rightarrow\infty$ or $u\rightarrow0$, respectively, as $s\rightarrow\infty$.  The constraint matching solution (solid teal curve) satisfies the boundary condition $\sigma(s=1\rm \,AU) = 0$.} }
    \label{Figure::iso_visc}
\end{figure}

Solutions to Eq. \eqref{Eqn::Parker} can be visualized by plotting its level sets for different values of $\mathcal{G}$ in an $M$--$s$ phase-space.
Representative solutions (for different values of $\mathcal{G}$) are shown in the black (dashed and dashed-dotted) curves in Figure \ref{Figure::iso_visc}.
{The bold dashed-dotted curves represent transonic solutions, which separate the supersonic wind solutions (above) from the subsonic breeze solutions (below).
The contours to the left and right of the transonic curves are nonphysical solutions that do not span the full spatial domain; 
however, portions of these solutions can be accessed through a shock transition from adjacent curves.

In contrast to Parker's analytical solution, the viscous isothermal equations take the form}
\begin{gather}
    \partial_s u = - \sigma / \mu \\
    \partial_s \rho = \rho \left ( \sigma / (\mu u) - \partial_s \ln A \right ) \\
    \partial_s \sigma = \dot{\mathcal{P}} + (\rho C_s^2 - \sigma) \partial_s \ln A + \frac{\rho \sigma}{\mu u} \left( u^2 - C_s^2 \right).
\end{gather}
These equations cannot be posed as a perfect differential, and must instead be {computed numerically} from initial conditions.
{The challenge, therefore, is to determine suitable initial values for $u$, $\rho$, and $\sigma$, based on reasonable boundary conditions in order that these equations can be solved as a well-constrained IVP.}

{
Our first task is to choose an appropriate boundary condition in order to constrain the solution at large $s$.
Normally this would entail the construction of an asymptotic solution in which the velocity tends to a constant value while the various fluxes decrease at least as quickly as $s^{-2}$, in order that the net global energy flux remain finite as $s\rightarrow \infty$.
As it happens, the viscous isothermal system admits no asymptotic solutions that obtain constant velocity at large distance.
Additionally, the familiar logarithmic solution, $M^2 \sim \ln A^2$, which describes Parker's isothermal wind at large distances, ceases to be a valid solution in the presence of viscous effects.
In fact, we have identified only two self-consistent asymptotic forms; breeze solutions, which are characterized by $u(s\rightarrow\infty)\propto s^{-1}$, and divergent wind solutions, for which $u(s\rightarrow\infty) \propto s^3$.
}

{
In the absence of a suitable asymptotic boundary condition, we have instead selected what we feel is a sensible intermediate choice, namely that the flow speed at $1\,\rm AU$ should be supersonic and constant.
The supersonic requirement is trivially satisfied for solutions with initial velocity similar to that of the Parker solution, while the constant speed at $1\,\rm AU$ reduces to a requirement on the viscous stress; namely $\sigma(s = 1\,\rm AU) \rightarrow 0^{-}$.
Since this constraint, which is fundamentally a Dirichlet condition on $\sigma$, targets only a single parameter, it should be possible to find a constraint satisfying solution by fixing two of our three initial conditions and then exploring a range of values for the third, casting the IVP as a shooting problem.
From our exploration of the solution space, we have found that the viscous isothermal solutions are only weakly sensitive to the initial values of $\rho$ and $\sigma$, while they depend quite strongly on the initial value of $u$.
We have, therefore, elected to specify $\rho_0$ and $\sigma_0$ from {a priori} considerations, while treating $u_0$ as a free parameter.}

{
For the initial density, we chose a reasonable value for the expected coronal condition: $\rho_0 = m_i \times 10^9\,\rm cm^{-3}$.
Then, to initialize $\sigma_0$ we considered two reasonable constructions, which are detailed in Appendix \ref{App::Init}. 
The first is a continuity argument, requiring consistency between $\partial_s u$ and the conditions of the atmosphere below $s=R_\odot$, while the second requires that $\divrm (\sigma)$ vanish at $s=R_\odot$, guaranteeing that the viscous and inviscid solutions agree at that point.
Each of these conditions gives $\sigma_0$ in terms of $u_0$, $\mu(T_0)$, and $L_u$, where $L_u$ is the characteristic length scale of $u$ as given in each case. 
Remarkably, both constructions give almost identical values of $L_u \simeq 2.5 \times 10^{10}\,\rm cm$, which, combined with $\mu(T_0) \simeq 2.95 \,\rm g\, cm^{-1}\, s^{-1}$, leads to a value of $\sigma_0 \approx - u_0 \times 10^{-10}\,\rm g\, cm^{-2}\, s^{-1}$.
This initial value is so small as to be indistinguishable from zero when compared to the other relevant terms and we have found that the computed solutions are insensitive the choice of $\sigma_0$ within the range $- u_0 \mu(T_0) / L_u < \sigma_0 < 0$.
For this calculation we have, therefore, elected to use the simplest initial condition that can be drawn from within this reasonable range: $\sigma_0 = 0$.
}

{
With $\rho_0$ and $\sigma_0$ set, the only remaining initial condition is $u_0$, which can be varied as necessary in order to identify a constraint satisfying solution.
Several possible solutions are depicted in Figure \ref{Figure::iso_visc}, including the divergent wind solutions and various other solutions that are subsonic at large distance.
The latter group can be further divided into simple breeze solutions, which are everywhere subsonic, and shocked breeze solutions, which exhibit a sonic point near the inviscid critical point but then decelerate through a region of strong viscous stress\footnote{Solutions of this type have been discussed previously by \citet{Axford:1967} and \citet{Owocki:1983}.} (i.e., a shock) before eventually settling to a subsonic breeze configuration, similar to the many subsonic contours of the Parker solution.}

{
The blue curve in Figure \ref{Figure::iso_visc} corresponds to the only solution for which $\sigma(s = 1\rm\,AU) = 0$, which results from the initial condition $u_0 = 4.48933905\times10^6\,\rm cm\, s^{-1}$.
This solutions lies between the families of divergent wind solutions and shocked breeze solutions at $1 \,\rm AU$.
It follows very closely to the transonic Parker solution, passing almost exactly through the Parker critical point, and only begins to diverge from the Parker solutions at $s\gtrsim10R_\odot$, beyond which it initially accelerates before the speed plateaus to $u(s=1\,\rm AU) \simeq 4 C_s$.
We should note, however, that this solution only matches the required boundary condition locally (at $s=1\,\rm AU$); were the calculation continued beyond that point for the same value of $u_0$ the solution would continue to decelerate, similar to a shocked breeze solution.
}

{
The constraint satisfying solution was found by first guessing an initial velocity that was informed by the Parker solution and then manually adjusting that initial guess until the constraint was satisfied.
Since the speed at $1 \,\rm AU$ increased monotonically with increases in $u_0$, this task was relatively straightforward, although the shape of the solution is quite sensitive to the initial value.
A manual search of the solution space was performed with about 1 hour of human effort, during which time roughly 100 individual solutions were computed, each of which required less than one second to complete.
}

\subsection{Viscous solutions with nonideal heating}\label{sub::num_cnd}

While the isothermal solution is an instructive example, our ultimate aim is to calculate solar wind profiles with realistic energy, including all of the physics described in Section \ref{Section::ODE}.
{When fully factored, the complete set of governing equations takes the form}
\begin{gather}
    \partial_s u = - \sigma / \mu; \label{EQN::ODE_du}\\
    \partial_s P = - \frac{f P}{\kappa T} + \frac{\sigma P}{\mu u} - P \partial_s \ln A; \label{EQN::ODE_dP}\\
    \partial_s \rho = \rho \left ( \frac{\sigma}{\mu u} - \partial_s \ln A \right ); \label{EQN::ODE_dr} \\
    \partial_s f = \dot{\mathcal{Q}} + \frac{u}{\gamma - 1}\frac{P f}{\kappa T} + \frac{\sigma(\sigma + P)}{\mu} - (P u + f)\partial_s \ln A; \label{EQN::ODE_df}\\
    \partial_s \sigma = \dot{\mathcal{P}} + \frac{P f}{\kappa T} + \frac{\sigma}{\mu u} (\rho u^2 - P) - (\sigma - P) \partial_s \ln A. \label{EQN::ODE_ds}
\end{gather}
{As with the previous example, we solve these equations as an IVP, integrating from initial conditions at the solar surface, subject to a set of imposed boundary conditions.}

\begin{figure*}[ht]
    \centering
    \includegraphics[width=\linewidth]{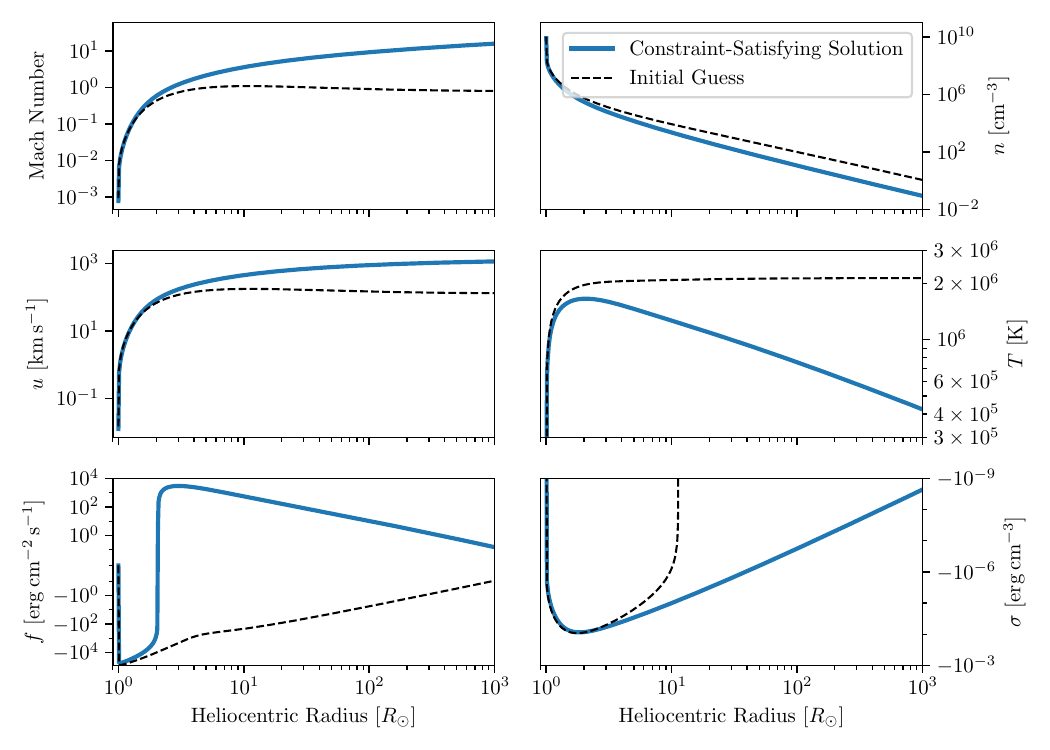}
    \caption{{Solution profiles of the complete set of viscous nonideal solar wind equations. 
    The black dashed curves are generated from a guess at the initial conditions, which is reasonable given the assumed mass flux of the solar wind, but which does not result in a constraint satisfying solution.
    The solid teal curves are the constraint satisfying solution, for which the initial conditions were systematically tuned (beginning with the initial guess) to obtain a solution that matches the asymptotic boundary condition.
    }}\label{Fig::Full_Sol}
\end{figure*}

{Unlike the viscous isothermal equations discussed in the previous section, equations \eqref{EQN::ODE_du} -- \eqref{EQN::ODE_ds} admit an asymptotic solution with finite energy flux and a supersonic outflow that tends to a constant value as $s\rightarrow\infty$.
This solution, which is derived in Appendix \ref{App::ASols} takes the form
\begin{align}
    f(s\gg R_\odot) &\rightarrow f_* \times (s/s_*)^{-2}, \notag \\
    \sigma(s\gg R_\odot) &\rightarrow \sigma_* \times (s/s_*)^{-2}, \notag \\
    T(s\gg R_\odot) &\rightarrow T_* \times (s/s_*)^{-2/7}, \notag \\
    u(s\gg R_\odot) &\rightarrow u_\infty + (u_* - u_\infty) \times (s/s_*)^{-2/7}.
\end{align}
The location $s_*$ is an arbitrary reference location, beyond which the solution attains its asymptotic form, while the subscript `$*$' indicates the values that the various primitive variables obtain at $s_*$.}

{As discussed in the Appendix, the assertion of this form, along with the constraints implied by the energy and momentum equations, results in a three-parameter solution space at $s_*$, meaning that if a solution were computed by integrating inward from $s_*$ toward $R_\odot$, it would be possible to satisfy three boundary conditions at $s=R_\odot$, while the remaining two values would be dictated by the solution.
Reversing this logic, it follows that if the same three boundary conditions are chosen as initial values when integrating outward from $R_\odot$ toward $s_*$, then a self-consistent asymptotic form can be obtained by tuning the remaining two parameters at $R_\odot$ until the asymptotic constraint is satisfied.
Put differently, if the asymptotic solution comprises two constraints (the power-law indexes of $f$ and $\sigma$) then there are three additional boundary conditions that can be imposed {a priori} at $R_\odot$, while the remaining two initial conditions at $R_\odot$ must be determined {a posteriori} (through exploration of the parameter space) in order for the solution to be self-consistent.}

{
We have chosen to set and hold fixed the initial conditions on $T$, $f$, and $\sigma$, while varying the initial conditions on $\rho$ and $u$, with $P$ given implicitly per equation \eqref{Eqn::IGas}.
For the temperature we choose $T_0 = 2\times10^4\,\rm K$, which is generally taken to represent the layer of the atmosphere at which the partially-ionized chromosphere meets the fully ionized corona at the base of the transition region.
As discussed previously, the viscous stress is initialized through consideration of the velocity length scale, $L_u$, and dynamic viscosity, $\mu(T_0)$.
Given the lower temperature at the base of the transition region relative to the corona, the relevant values are $\mu(T_0) \sim 9.5\times10^{-6}\rm \, g\, cm^{-1}\, s^{-1}$ and $L_u \sim 9.3 \times 10^7\,\rm cm$ (computed, as before, from arguments given in Appendix \ref{App::Init}).
From these we then obtain $\sigma_0 = - \mu(T_0) u_0 / L_u \simeq - u_0 \times 10^{-13}\,\rm g\, cm^{-2}\, s^{-1}$.
As with the viscous isothermal calculation, we have found this value to be indistinguishable from $\sigma_0 = 0$, which is the value that we impose.}

{
The heat flux at the base of the transition region has been discussed extensively by \citet{Gonczi:1977} and \citet{Hammer:1982}, who argued that the radiative losses at the base of the transition region require
\begin{equation}
    f_0^2 < |\mathcal{R}_0 \kappa(T_0) T_0|.
\end{equation}
While the upper limit of this range may be more realistic (one expects a nonzero downward conductive flux from the corona into the chromosphere), the resulting solution appears to be insensitive to any choice of $f_0$ satisfying this constraint.
Following the precedent of previous authors \citep[see, e.g.,][]{RTV:1978} we have (again, for simplicity) elected to enforce the lower limit of this condition: $f_0 = 0$.}

\begin{figure*}[h!]
    \includegraphics[width=\linewidth]{fig3.pdf}
    \caption{Asymptotic profiles of the constraint satisfying solution plotted with respect to heliocentric radius ($s$). Top panel: Profiles of $T/T_*$, $(u_\infty - u ) / (u_\infty - u_*)$, and an $s^{-2/7}$ power law for reference. Middle panel: Profiles of  $f/f_*$, and $\sigma/\sigma_*$, with an $s^{-2}$ power law for reference. In both cases, the computed solutions are in good agreement with the representative power laws in the asymptotic region ($r \gtrsim 10^4 R_\odot$). Bottom panel: Total energy flux associated with heat conduction ($f$), viscous stress ($\sigma u$), bulk kinetic energy ($\rho u^3 / 2$), as well as saturation limits associated with free-streaming of electrons ($p_e v_e$) and ion thermal energy density ($p_i u$). The total energy flux, which is proportional to the sum of ($\rho u^3 / 2 + f + \sigma u$), is shown in the black dashed-dotted curve.}\label{Fig::ASol}
\end{figure*}

{With $T_0$, $f_0$, and $\sigma_0$ fixed, it remains to explore the space of (reasonable) values for $\rho_0$ and $u_0$ to determine an initial condition that generates a constraint satisfying solution, consistent with the asymptotic form discussed above.
Our approach was to compute solutions out to a distance of $s=10^5R_\odot$, well beyond the point where the asymptotic assumptions should be satisfied, and then to compare the profiles of $f$, $\sigma$, $T$, and $u$, to the expected power-law profiles.
For this we chose an initial guess of $u_0=1.5\times10^{3}\rm\, cm\, s^{-1}$ and $n_0 = 10^{10}\, \rm cm^{-3}$ (with $\rho_0 = m_i n_0$), which we determined from the expected density at the base of the transition region and a crude estimate of the expected mass flux at $1\, \rm AU$.
The values of $u_0$ and $n_0$ were then adjusted manually until the computed solution satisfied the asymptotic constraints.}

{
Each realization of the calculation required approximately $1\,\rm second$ to complete, while the search required several hundred realizations to find the best solution, which amounted to about $2\,\rm hours$ of human effort.
The constraint satisfying solution found to result from initial conditions $n_0 = 8.0239133 \times 10^9\,\rm cm^{-1}$ and $u_0 = 1.2629299\times 10^3\,\rm cm\, s^{-1}$. 
The resulting profile is depicted by the solid teal curves in Figure \ref{Fig::Full_Sol}, which show the spatial profiles of the primitive variables (replacing $P$ with $T$) and the Mach number as functions of $s$.
The solution that results from the initial guess is also shown in the dashed black curves.
}

{
Both sets of initial conditions result in solutions that exhibit a transition region, through which the temperature rises abruptly to coronal values ($\sim 10^6\,\rm K$) while the number density ($n$) decreases abruptly from $\sim 10^{10}\,\rm cm^{-3}$ to a $\sim 10^8\,\rm cm^{-3}$ over a distance of less than $10\,\rm Mm$.
Both solutions exhibit speeds in excess of $100\,\rm km\, s^{-1}$;
however, the solution that results from the initial guess exhibits a monotonic increase in $T$ with height, which generates a conductive flux that is everywhere negative (inward) and leads an outflow that is subsonic at large $s$.
By comparison, the constraint satisfying solution has a local temperature maximum of $\sim 2\times10^6\,\rm K$, which occurs at a height of $s\sim2 R_\odot$, beyond which the temperature decreases with height, resulting in a positive (outward) heat flux at large $s$. 
The reduction in temperature with height also causes the Mach number to increase steadily with increasing $s$, even as $u$ approaches an asymptote.

The fluxes associated with the constraint satisfying solution are shown in Figure \ref{Fig::ASol}, along with asymptotic profiles of $T$, $u$, $f$, and $\sigma$.
The asymptotic forms appear to be satisfied to good accuracy for all $s\gtrsim10^4\,R_\odot$, which is the value that we chose for $s_*$ when enforcing the asymptotic constraint.
The global energy fluxes associated with conduction and viscosity (which is negative in the corona and beyond), are shown in the bottom panel, along with the relevant saturation limits ($p_e v_e$ in the case of electron heat flux and $p_i u$ for the ion-dominated viscous stress).
Here $p_{i,e} = P/2$ are the ion and electron partial pressures, and $v_e = \sqrt{k_B T / m_e}$ is the electron thermal speed.
The net (global) energy flux, shown in the dashed-dotted black curve, is dominated by conductive, viscous, and mechanical energy (the enthalpy contribution being negligible), and quickly obtains an asymptotic value of $\sim 2\times10^{27}\,\rm erg\, s^{-1}.$
}

{
The salient features of the constraint satisfying solution are these:
The fluxes of conductive, viscous, and mechanical energy all scale as $s^{-2}$ at large $s$, meaning that each of these generates a finite but nonzero total energy flux when integrated over a solid angle of $4\pi\,\rm steradians$.
The total combined energy flux comprises the sum of conductive, viscous, and mechanical contributions, and is also finite and nonzero as $s\rightarrow0$; 
although it is smaller than the sum of conductive and mechanical energy fluxes, owing to the inward energy flux associated with the viscous stress.
The speed is asymptotic to a constant value of $u_\infty \sim 1.5\times10^3\,\rm km\, s^{-1}$ at large $s$, and the temperature decreases as $s^{-2/7}$ from a maximum of $2\,\rm MK$ in the lower corona.
At $s=1\,\rm AU\, (\sim 200\,R_\odot)$ this solution predicts a wind speed of $ \sim 10^3\rm \, km\, s^{-1}$, particle densities of $\sim 2.5\,\rm cm^{-3}$, and a temperature of $\sim 6 \times10^5\,\rm K$, all of which are reasonable for the fast solar wind, albeit slightly in excess of observed values \citep[see, e.g.,][]{Verscharen:2019}.
}


\section{Discussion}\label{Section::Discussion}

The introduction of the viscous stress through the closure in Eq. \eqref{Eqn::VscClsr} provides a remedy for the singular condition that arises when the inviscid hydrodynamic equations that govern the solar wind are formulated as an initial value problem.
The resulting system of {stiff} ODEs can be integrated numerically without consideration of the presence of a critical point; 
however, the resulting solution is sensitive to the initial conditions, which must be chosen such that the solution satisfies {appropriately constructed asymptotic boundary conditions}.
Nonetheless, for suitable initial conditions the solutions generated by this method are robust, self consistent, and can be calculated in just a few seconds -- orders of magnitude faster than methods that depend on the relaxation of the time-dependent Navier-Stokes equations.

The example solution depicted in Fig. \ref{Fig::Full_Sol} has many of the features that we require of a realistic solar wind solution,
{including a transition region, where the plasma accelerates rapidly while the density decreases and the temperature jumps from chromospheric to coronal values.}
It also exhibits an inward conductive heating flux in the lower corona, which is necessary to support the strong radiative cooling there, below an extended middle and outer corona in which the temperature decreases monotonically with distance from the sun. 
Also, and importantly, the conductive heat flux and viscous stress both {decrease with the square of the heliocentric distance, meaning that they generate finite but appreciable total energy fluxes.}

{Solutions of this kind have been computed previously, as discussed by \citet{Summers:1980} and others, but never with the inclusion of realistic heating and cooling in the lower corona, and never from initial conditions at the base of the corona.
In that sense, the calculations presented here represent a first-of-their-kind attempt to extrapolate realistic solar wind profiles from reasonable conditions near the solar surface, through the transition region and lower corona, and into the extended heliosphere, while matching physically motivated asymptotic boundary conditions.
The fact that solutions of this kind can be computed with relatively little computational cost is a testament to the rapid increase in available computing resources over recent decades and the long history of improvements in numerical methods since these equations were first discussed in the mid 1960s.
}

This model also has some obvious limitations.
{The most notable of these are the magnitudes of the conductive heat flux and viscous stress in the tenuous outer corona and heliosphere, where they substantially exceed the saturation limits associated with the energy flux of free-streaming electrons and ion enthalpy flux.}
The likeliest causes of this overestimate are the closures that {that we have employed, which are based on the collisional models of \citet{Spitzer:1953} and \citet{Braginskii:1965}, and which are known to} admit arbitrarily large values for $f$ and $\sigma$.
This condition can likely be alleviated through the introduction of flux limiters, and we have suggested {two possible approaches} in Appendix \ref{App::FLims}.

{
The transition from collisional to collisionless regimes, and the appropriate closures of the conductive heat flux have been discussed extensively by authors such as \citet{Bale:2013} and \citet{Landi:2012}.
\citet{Cranmer:2021} have also investigated closure relationships based on specific parameterizations of non-Maxwellian velocity distribution functions, and find that these too have difficulty matching observed energy fluxes.
More generally, the implications of closure statements that lead to self-defeating conditions have also been discussed at length by \citet{Scudder:2019} -- in summary, the task of constructing self consistent closures, whose underlying assumptions are satisfied within the calculations that evoke them, is exceedingly difficult.
For the purposes of the present discussion we have elected to avoid regularization in favor of mathematical clarity; however for this model to be deployed for forecasting purposes it will be necessary to validate its predictions and employ a more sophisticated closure scheme.}

Another limitation of this model is the time required to identify suitable initial conditions given the specified constraints on the asymptotic solution properties.
For this example, we specified the temperature, heat flux, and viscous stress based on physical {considerations} and then performed a manual search over the remaining primitives in order to find initial conditions that generated a constraint satisfying solution, consistent with asymptotic constraints.
In order for this method to be deployed at scale, this parameter search will need to be automated, and we are currently exploring various techniques for performing this task.
{Based on preliminary work we estimate that an automated search of this kind would require several hundred or perhaps as many as a thousand individual solutions to be calculated. 
This search will undoubtedly introduce additional computational cost; 
however, even if the compute required to identify a constraint satisfying solution proves to be $10^3$ times the cost of an individual calculation, the total cost is expected to be not more than one CPH-hour -- far less than the time required for a time-dependent simulation to relax to a steady state.}


\section{Conclusion}\label{Section::Conclusion}

The singular conditions that emerge at the sonic point in conventional, inviscid models of the solar wind {require complex solution algorithms}, which are an impediment to the determination of steady-state solar wind profiles as an initial value problem.
{
Viscous models are known to be free of such singularities; however, these are more computationally demanding and require more sophisticated boundary conditions in order to avoid spurious results.
}
Consequently, the extrapolation of the speed, density, and temperature of the solar wind from the solar surface into the heliosphere has generally not been {considered practical}.
By (re)introducing the viscous stress through the Newtonian closure {and applying modern numerical methods (see Appendix \ref{App::num})}, we have remedied {these concerns} and demonstrated that realistic solar wind profiles, complete with external heating, radiative cooling, and thermal conduction, can be calculated directly from the steady-state hydrodynamic equations.

{
The ability to determine steady-state solar wind profiles by direct integration will remove a considerable computational bottleneck for both global heliospheric modeling and time-dependent numerical experiments, which are typically initiated from steady-state conditions that must be computed in advance \citep[see, e.g.,][]{Scott:2022a, Scott:2024}.
Once this method has been refined and bench-marked against comparable time-dependent models, it promises to deliver a considerable increase in computation performance over the current state of the art.}
By populating the plasma parameters along a defined magnetic field line based on a first principles calculation, which can be performed in a matter of minutes on a desktop computer, with realistic heating and cooling, we {expect to} able to construct global heliospheric models of the solar wind in near-real-time.
Moreover, initial conditions for time-dependent simulations will incur considerably less computational cost, meaning that numerical experiments can be set up and run with substantially less effort than is currently required.

The method that we have presented is only an initial proof of concept of this technique, but the calculations presented here are encouraging.
As we refine this technique {we anticipate that the fidelity of these solutions} will only improve, allowing us to explore the effects of different heating models and magnetic geometries at considerably lower computational cost than has previously been the case.
We are optimistic that this technique, and the improved models that follow from it, will be instrumental in shaping the next generation of solar and heliospheric wind models.


\begin{acknowledgements}
This work was supported by the Office of Naval Research 6.1 Program, NASA PSP/WISPR grant NNG11EK11I, and by a NASA Heliophysics Supporting Research grant under ROSES NNH20ZDA001N ``Investigating the Influence of Coronal Magnetic Geometry on the Acceleration of the Solar Wind'' (Science PI: RBS). {The authors are indebted to the anonymous referee for their careful reading and excellent feedback on the manuscript. These contributions have led to a substantial improvement in the quality of the work that we have presented here. }
\end{acknowledgements}


\bibliographystyle{aa}
 \newcommand{\noop}[1]{}

\appendix 

\section{Numerics}\label{App::num}

{
In order to integrate the various systems of ODEs discussed in Section \ref{Section::ODE} from initial conditions at the solar surface, we have developed a set of computational tools that we call HelioSHOT, which have been published in the form of a Jupyter Notebook of the same name \citep{HelioSHOT_zenodo}.
The code is written in Python and relies on standard Python libraries, including the Numba JIT (Just in Time) compilation library, which compiles the various functions into C code for faster execution.
The notebook contains both the integration tools and a selection of custom function definitions for stating the particular form of the problem being solved in each case.
The routines have been written to be agnostic to the particular equations being solved, but we have chosen the integration scheme to be suitable to {stiff} systems of ODEs, which are problematic for explicit solvers.}

{
HelioSHOT's core integration routine consists of an implicit trapezoid solver that uses fixed-point iteration (FPI) to find the update to the state vector in order that the finite difference  estimate of the derivative, as evaluated over a given spatial interval, is equal to the average of the analytical gradient when evaluated at the two end points of the interval. 
That is, if $\mbf{U}(s)$ is the state vector at a point $s$, and if $\mbf{G}(\mbf{U}; s) \equiv \partial_s \mbf{U}(s)$ is the gradient of the state vector as defined according to the system of ODEs, then the implicit trapezoid scheme requires that
\begin{equation}
    \mbf{U}(s_+) - \mbf{U}(s_-) = \frac{\Delta}{2} \left ( \mbf{G}(\mbf{U}(s_+); s_+) + \mbf{G}(\mbf{U}(s_-); s_-) \right ),
\end{equation}
where $s_-$ and $s_+$ correspond to the endpoints of the spatial interval and $\Delta = s_+ - s_-$ is spatial extent (i.e., the step size).
Since $\mbf{U}(s_+)$ is not initially known, and $\mbf{G}(\mbf{U}; s)$ is not directly invertible, this implicit scheme requires a nonlinear solver to find $\mbf{U}(s_+)$.
The FPI algorithm begins with an explicit Euler step as a first guess of the updated state vector;
\begin{equation}
    \mbf{U}^{(0)} = \mbf{U}(s_-) + \Delta \mbf{G}(\mbf{U}(s_-);s_-),
\end{equation}
and then iteratively updates the guess by comparison to the trapezoid rule:
\begin{align}
    & \bar{\mbf{U}} = \mbf{U}(s_-) + \frac{\Delta}{2} \left ( \mbf{G}(\mbf{U}^{(i)}; s_+) + \mbf{G}(\mbf{U}(s_-); s_-) \right );\\
    & \mbf{U}^{(i+1)} = W  \bar{\mbf{U}} + (1-W) \mbf{U}^{(i)},
\end{align}
where $W(=0.5)$ is a safety factor to prevent overshooting. 
This iterative update proceeds until a tolerance (t) is reached so that 
\begin{equation}
   t^{(i)}\left (  \frac{\mbf{U}^{(i+1)} - \mbf{U}^{(i)}}{\mbf{U}^{(i+1)} + \mbf{U}^{(i)}} \right )^2< t,
\end{equation}
at which point the converged value is taken as a solution to the implicit step: $t_i < t \rightarrow U(s_+) \equiv U^{(i+1)}$.

This approach has the benefit of being reversible up to the accuracy of the fixed-point solver, meaning that the solutions should not change depending on whether they are computed in the direction of increasing or decreasing $s$ (a property that is not exhibited by explicit schemes).
The implicit trapezoid scheme is also known to perform well on {stiff} systems of ODEs, especially when combined with adaptive stepping, which typically requires some built-in error estimation for the control logic.
In this case, however, we have used the fixed point iteration itself to control the adaptive stepping directly by requiring that the number of iterations required for the FPI to converge--which serves as a proxy for the error estimation--fall within a specified range. 
If the solution converges too quickly (less than two iterations) then a larger step can be taken, while slow convergence (greater than $~20$ iterations) leads to a reduction of the step and recalculation of the update.

With HelioSHOT's exponentially spaced grid and adaptive stepping it is able to take extremely small steps near the solar surface ($\Delta s \lesssim 5\,\rm m$) while relaxing the step size as $s$ becomes large, and since the step size is connected to the FPI routine, there is minimal computational overhead from sub-cycling of the FPI.
This is a considerable benefit to the efficiency of the solver and substantially reduces the time required to compute individual solutions for given specification of the initial conditions and model parameters.
For a typical calculation of the viscous conducting model the compute time to return a solution that spans from $R_\odot \leq s \leq 10 \rm \, AU $ is about $0.3\,\rm seconds$, while increasing the domain to $R_\odot \leq s \leq 500\,\rm AU$ only increases the compute time to $\sim 0.8\,\rm seconds$. 

More details of the code and its numerics are available in the source documentation \citep[see][]{HelioSHOT_zenodo}.
}

\section{Initialization of $f$ and $\sigma$}\label{App::Init}

{
According to \citet{Gonczi:1977}, the structure of the transition region is such that there is a boundary layer through which the conductive heat flux changes more rapidly than the temperature or density, with the consequence that the solution is insensitive to the particular value of $f$ below this layer. 
\citet{Hammer:1982} characterized the threshold values for $f$ at the base of the transition region and found that the structure of the coronal solution was insensitive to any value of $f_0$ in the range
\begin{equation}
-\sqrt{|\mathcal{R}_0 \kappa(T_0) T_0|} < f_0 < 0 
\end{equation}
where $\mathcal{R}_0$ is the radiative cooling rate at the base of the transition region.
A similar argument was made by \citet{RTV:1978} in their construction of the loop atmosphere scaling laws.
}

{
While the coronal solution is generally unaffected by the choice of $f_0$ within this range, it has been suggested that the most realistic choice corresponds to $f_0^2 \lesssim |\mathcal{R}_0 \kappa(T_0) T_0|$, since it is expected that some amount of heat energy should be conducted through the transition region and into the underlying chromosphere; however, the argument for this is primarily an intuitive one since the models discussed here are not likely to provide a robust description of the transport processes that mediate the energy exchange between the fully ionized corona and the partially ionized chromosphere.
We submit, therefore, that for nonkinetic models assuming local thermodynamic equilibrium, $f_0=0$ is as reasonable an estimate as any within this range.
}

{
In order to initialize the viscous stress at the base of the transition region we considered two practically motivated constraints:
The first is the requirement that $\divrm (\sigma) = 0$ at $s=R_\odot$, which ensures that there is no viscous force on the fluid at the bottom boundary. 
This constraint, which guarantees that the momentum balance of the viscous and inviscid solutions are identical at $s=R_\odot$, is achieved by solving the inviscid momentum equation for $\partial_s u$ and then constructing $\sigma$ from the closure equation \eqref{Eqn::VscClsr}, or, equivalently, by removing $\partial_s \sigma$ and $\sigma \partial_s \ln A$ from Eq. \eqref{EQN::ODE_ds} and solving for $\sigma$ as a function of the other primitive variables.
The resulting initial condition for $\sigma$ is
\begin{equation}
    \sigma_0 = -\mu(T_0) u_0 \times \left ( \frac{2 P_0 / R_\odot - \rho_0 g_\odot}{\rho_0 u_0^2 - P_0} \right ),
\end{equation}
where we have assumed spherical expansion and written $\partial_s \ln A = 2/R_\odot$.

The second scenario assumes that the atmosphere below the lower boundary is effectively hydrostatic with only a weak outflow that is far slower than the sound speed, and then requires that the velocity gradient at $s=R_\odot$ is consistent with the density gradient of this underlying hydrostatic atmosphere under the continuity equation.
If the underlying atmosphere is isothermal, then the local density gradient satisfies $\partial_s \ln \rho(s=R_\odot) = L_\rho^{-1}$, where $L_\rho = P_0 / (\rho_0 g_\odot)$ is the density scale height at $s=R_\odot$.
The velocity gradient is then found from the continuity equation, again with $\partial_s \ln A = 2/R_\odot$, and the corresponding value of $\sigma_0$ is again constructed from $\mu(T_0)$ and $\partial_s u(R_\odot)$.
This leads to the initial condition
\begin{equation}
    \sigma_0 = -\mu(T_0) u_0 \times \left ( \rho_0 g_\odot / P_0 - 2 / R_\odot \right ).
\end{equation}
Interestingly, this value is identical to the previous expression in the limit $\rho u^2 \ll P$, and is approximated to good accuracy by
\begin{equation}
    \sigma_0 \simeq - \mu(T_0) u_0 \times \rho_0 g_\odot / P_0,
\end{equation}
since $L_\rho \ll R_\odot$.
}

\section{Asymptotic solution} \label{App::ASols}

{
In lieu of specifying a local boundary condition at a particular point in space, we seek, instead, an asymptotic solution to the Navier-Stokes equations \eqref{Eqn::ConsContinuity} -- \eqref{Eqn::ConsEnergy}, with the particular closures of $f$ and $\sigma$ given in equations \eqref{Eqn::HfxClsr} and \eqref{Eqn::VscClsr}, and in the absence of any external heating, cooling or forcing other than gravity (all of which are assumed to be negligible at large distance). 
Note that for the viscous isothermal limit discussed in Section \ref{sub::vsc_iso} we have identified no asymptotic solutions that tend to a constant nonzero velocity at large distances from the Sun; however, for the nonideal viscous equations with thermal conduction there does appear to be a suitable asymptotic solution.
}

{In order to construct such a solution we first assume that $T$ and $u$ take the form
\begin{align}
    T(\lambda) & = T_* \times \lambda^\alpha, \quad \text{and}\\
    u(\lambda) & = u_\infty \times (1 - \delta_*\lambda^{\beta}),
\end{align}
where $\lambda = s_* / s$ is our independent variable and the expansion is spherically symmetric so that $A(s) \propto s^2$.
The reference location $s_*$ denotes an arbitrary point beyond which the asymptotic solution obtains, and the `$*$' subscript indicates the value that each primitive (dependent) variable attains at $s=s_*$ ($\lambda = 1$).
The limit $s/s_* \rightarrow \infty$ corresponds to $\lambda \rightarrow 0$, and we require that $T(\lambda \rightarrow 0) \rightarrow 0$ and $u(\lambda \rightarrow 0) \rightarrow u_\infty$, while $T(\lambda=1) = T_*$ and $u(\lambda=1) = u_* = u_\infty \times (1-\delta_*)$.
The conductive heat flux and viscous stress then take the form
\begin{align}
    f(\lambda) & = f_* \times \lambda^{\frac{7\alpha}{2} + 1}, \quad \text{and}\\
    s(\lambda) & = s_* \times \lambda^{\frac{5\alpha}{2} + \beta + 1},
\end{align}
where $f_* = \alpha \kappa_* T_*/ s_*$ and $\sigma_* = - \beta \mu_* \delta u_\infty / s_*$ as required by the closure relationships.
}

{
If we impose the least restrictive condition on the total energy flux per unit solid angle (sr) that is consistent with the assumption of finite energy, then it suffices to require that $(f + \sigma u) \propto \lambda^2$.
This is achieved by setting $\alpha = \beta = 2/7$, from which it follows that
\begin{align}
    f(\lambda) & = f_* \times \lambda^2, \\
    \sigma(\lambda) & = \sigma_* \times \lambda^2;
\end{align}
while
\begin{align}
    T(\lambda) & = T_* \times \lambda^{2/7},\\
    u(\lambda) & = u_\infty \times ( 1 - \delta_* \lambda^{2/7} ).
\end{align}
It then follows from the continuity equation and the ideal gas law that
\begin{align}
    \rho(\lambda) &= \rho_* \lambda^2 u_* / u(\lambda),\\
    P(\lambda) & = P_* \times \lambda^{2+2/7} u_* / u(\lambda).
\end{align}
Provided that we can demonstrate the validity of such a solution, it will exhibit finite but nonzero net fluxes of viscous, kinetic, and conductive heat energy, while the enthalpy flux ($ F_P \propto r^2 P u \propto \lambda^{2/7}$) will be negligible as $\lambda \rightarrow 0$.
}

{To see that these forms constitute an asymptotic solution to the Navier-Stokes equations, we need only evaluate equations \eqref{EQN::ODE_df} and \eqref{EQN::ODE_ds} with the proposed asymptotic solution, since the continuity equation and the closures of $f$ and $\sigma$ have been satisfied by construction.
By matching coefficients to the lowest order in $\lambda$, we can then confirm the validity of the solution, since higher order terms will vanish as $\lambda \rightarrow 0$.
The resulting relationships between the leading coefficients then constrain the various dimensionless quantities at the reference location ($s_*$), which must satisfy the given criteria in order for the asymptotic solution to obtain.
}

{
It is convenient to begin with equation \eqref{EQN::ODE_ds}, and to note that $\divrm(\sigma)=0$, from which it follows that
\begin{align}
        0 & = \rho g(\lambda) + \frac{P f}{\kappa T} + \frac{ \rho u \sigma}{\mu} - \frac{P \sigma}{\mu u} + P \partial_s \ln A.
\end{align}
After replacing the primitive variables with their asymptotic equivalents and noting that $\partial_s \ln A = 2/s = 2 \lambda / s_*$, we then have
\begin{align}
    0 & = \left( \left(\frac{P_* f_*}{\kappa_* T_*} + \frac{2 P_*}{s_*}\right) + \frac{\rho_* u_* \sigma_*}{(1 - \delta_*) \mu_*} \right )\lambda^{3+2/7} \notag\\
    & + \frac{P_* f_*}{\kappa_* T_*} (1 + \delta_* \lambda^{2/7} + ...) \delta_* \lambda^{3+4/7} \notag\\ 
    & - \frac{P_* \sigma_*}{\mu_* u_*} (1 - \delta_*)( 1 + \delta_* \lambda^{2/7} + ...)^2 \lambda^{3+4/7} \notag\\
    & + \frac{2 P_*}{s_*} (1 + \delta_* \lambda^{2/7} + ...) \delta_* \lambda^{3 + 4/7} \notag\\
    & + \rho_* g_* \lambda^4 ( 1 + \delta_* \lambda^{2/7}),
\end{align}
where we have made use of the binomial expansion to write $u_* / u \simeq (1 - \delta_*) (1 + \delta_* \lambda^{2/7} + \mathcal{O} (\lambda^{4/7}))$ and discarded the higher-order terms.
Since there are multiple terms of order $\lambda^{3+2/7}$, this expression can be satisfied to arbitrary precision as long as the leading coefficients sum to zero.
}

{To leading order, this requirement becomes
\begin{align}
    0 & = (1 - \delta_*) \left(\frac{P_* f_*}{\kappa_* T_*} + \frac{2 P_*}{s_*}\right)  + \frac{\rho_* u_* \sigma_*}{\mu_*} \notag \\
    & = (1 - \delta_*) \left(\frac{2}{7}\frac{P_*}{s_* } + \frac{2 P_*}{s_*}\right)  - \frac{2}{7} \frac{\rho_* u_* \delta_* u_\infty }{s_*}
\end{align}
from which it follows that
\begin{equation}\label{Eqn::delta_M_constraint}
    M_*^2 = \frac{8 (1-\delta_*)^2}{\delta_*} = \frac{8 u_*}{u_\infty - u_*},
\end{equation}
where $M_*^2 = u_*^2 / c_*^2$ is the Mach number at $s_*$, $c_*^2 = P_* / \rho_*$ is the isothermal sound speed at $s_*$, and we have made use of the closure constraints on $f_*$ and $u_*$ given above.
Hence, the asymptotic form of the momentum equation, which is dominated by the acceleration of the fluid in response to the pressure force, constrains the asymptotic Mach number to take a particular form based on the relationship between $u_*$ and $u_\infty$ (which are related through $\delta_*$).
}

{
We can, similarly, inspect the ODE representation of the energy equation \eqref{EQN::ODE_df}, with $\divrm (f) = 0$, which then becomes
\begin{align}
    0 & = \frac{u}{\gamma - 1}\frac{P f}{\kappa T} + \frac{\sigma(\sigma + P)}{\mu} - P u \partial_s \ln A.
\end{align}
After replacing the primitive variables with their asymptotic forms, and making the same substitution for $\partial_s \ln A$, we then have
\begin{align}
   0 = & \left ( \frac{1}{\gamma-1}\frac{f_* P_* u_*}{\kappa_*T_*} - \frac{2 P_* u_*}{s_*} + \frac{\sigma_*^2}{\mu_*} \right ) \lambda^{3+2/7} \notag \\ 
   & + (1 - \delta_*) \frac{P_* \sigma_*}{\mu_*} \left ( \lambda^{3+4/7} + \delta_* \lambda^{3+6/7} + ... \right ).
\end{align}
As before, this expression can be satisfied to order $\lambda^{3 + 2/7}$, which is the scale at which viscous heating balances adiabatic cooling.

In order to enforce energy conservation we gather terms at lowest order and substitute $f_*$ and $\sigma_*$ from above, after which we have
\begin{equation}
    \left ( \frac{1}{\gamma-1} - 7\right ) \frac{P_* u_*}{s_*} + \frac{2}{7}\frac{\delta_*^2}{(1 - \delta_*)^2} \frac{\mu_* u_*^2}{s_*^2} = 0.
\end{equation}
This condition can be stated in terms of the Reynolds number at $s_*$, which is constrained to be
\begin{equation}
    R_e^* = R_e(\lambda=1) = \frac{\mu_*}{\rho_* u_* L_*} = \left(\frac{7\gamma - 8}{\gamma-1}\right) \frac{(1-\delta_*)^2}{\delta_*^2 M_*^2},
\end{equation}
where the length scale is taken to be $L_* = 7\, s_* / 2$. 
Then, since $M_*$ already depends directly on $\delta_*$, it follows that  
\begin{equation}\label{Eqn::delta_R_constraint}
    \delta_* \times R_e^* = \frac{7 \gamma - 1}{8 \gamma - 1} \simeq 0.865.
\end{equation}
}

{
It is surprising at first that the Prandtl number, given by
\begin{equation}
    P_r^* = \frac{\gamma}{\gamma - 1} \frac{k_B}{m_i} \frac{\mu_*}{\kappa_*},
\end{equation}
does not appear to be directly constrained by this solution.
This is a consequence of the fact that the conductive heating rate, $-\divrm(f)$, vanishes identically for all $\lambda$, meaning that $f_*$ does not contribute to the asymptotic energy budget.
Consequently, the asymptotic solution is independent of $\kappa_*$, whereas it depends directly on $\mu_*$, and so $R_e^*$, $M_*$, and $\delta_*$ are mutually constrained while $P_r^*$ is not. 
For practical purposes, however, any specification of $T_*$ automatically implies $\kappa(T_*)$ and $\partial_s T_*$, and is, therefore, equivalent to the specification of $f_*$.}

{
Since the full solution at $s=s_*$ is described by five parameters, enforcement of the asymptotic form (as constrained by the conservation of energy and momentum) reduces the number of free parameters that can be self-consistently specified, without violating the asymptotic constraint, to three.
Therefore, if (for example) $u_*$ and $T_*$ are known, along with the reference location $s_*$,  
then $f_*$ and $M_*$ are known automatically, while $\rho_*$, $\delta_*$ are given by equations \eqref{Eqn::delta_M_constraint} and \eqref{Eqn::delta_R_constraint}.
From $u_*$ and $\delta_*$, we then also have $u_\infty$ and $\sigma_*$, which completes the specification of the solution.
It is not important which parameters are specified; $\rho_*$, $u_*$, and $P_*$ could also be chosen, and similar logic would lead to the determination of $f_*$ and $\sigma_*$.
The critical point is that the specification of any three parameters, as well as enforcement of the asymptotic form, amounts to specification of the entire solution.
}


\section{Flux limiters}\label{App::FLims}

The conductive heat flux and viscous stress that we have included in our model are not bound by the physical constraints of the free-streaming electron heat flux or the diagonal components of the viscous stress tensor (i.e., the pressure). 
{
Therefore, while these closures may be justified in the lower portions of the solution (near the base of the corona) it is unlikely that they will lead to solutions with completely justified closures \citep[CJCs,][]{Scudder:2019}. 
In order to reconcile this, we suggest two alternate closure strategies that may be employed in future work.

First, it is possible to construct alternate formulations of the coefficients of viscosity and thermal conduction that are valid (or, at least, better justified) in the collisionless regime. 
For example, \citet{Williams:1995} derived transport coefficients on the assumption that the velocity distribution function is isotropized on the length scale of the gyro-radius, as a consequence of wave-particle interactions with a turbulent magnetic field.
This leads to the following definitions of the dynamic viscosity and thermal conductivity:
\begin{equation}
    \tilde{\mu}_{(i,e)} \propto \frac{m_{(i,e)}\, c}{q B} \times n_{(i,e)} k_B T_{(i,e)}
\end{equation}
\begin{equation}
    \tilde{\kappa}_{(i,e)} \propto \frac{5}{2}\frac{k_B}{m_{(i,e)}} \times \tilde{\mu}_{(i,e)},
\end{equation}
where $c$ is the speed of light, $q$ is the fundamental electric charge, and $B$ is the average local magnetic field strength.
These values are typically $\sim10^6$ times smaller than the usual Spitzer-H\"arm values, which should result in significant reductions in $f$ and $\sigma$.

Alternatively, it is possible to frame the closures using the Spitzer-H\"arm coefficients but with explicit regularizations that prevent $f$ and $\sigma$ from exceeding the saturation limits.
This can be done by inverting the usual algebraic regularization in terms of the harmonic mean to construct modified versions of $f$ and $\sigma$, from which $\partial_s T$ and $\partial_s u$ can then be found.
For example, if $f_\kappa$ is the conductive heat flux from Fourier's law, and $f_{v_e}$ is the free-streaming limit, then} the regularized heat flux can be defined \citep{Cowie:1977} as
\begin{equation}
    \bar{f} \equiv f_\kappa f_{v_e} / \sqrt{f_\kappa^2 + f_{v_e}^2},
\end{equation}
which is everywhere parallel to $f_\kappa$, and tends to $f_\kappa$ when $f_{v_e}$ is large, but is bounded by $f_{v_e}$ when $f_\kappa$ is large.
In order to include the free streaming limit, we can use $\bar{f}$ as a primitive variable, which specifies the conductive heating rate and is integrated in exactly the same way as $f_\kappa$ from its contribution to the energy equation, and then define the modified closure relationship
\begin{equation}
    \kappa(T) \partial_s T = -f_{v_e} \bar{f} / \sqrt{f_{v_e}^2 - \bar{f}^2},
\end{equation} 
which is used to specify $\partial_s T$ and, hence, $\partial_s P$.
This regularization will be well-behaved so long as that ${\bar{f}}^2 < f_{v_e}^2$, which can be enforced during numerical integration.

Similarly, for the viscosity, it is desirable to define the regularization of the viscous stress as
\begin{equation}
    \bar{\sigma} \equiv \sigma P / \sqrt{\sigma^2 + P^2}.
\end{equation}
This is necessary to prevent the off-diagonal terms in the viscous stress tensor from exceeding the trace (which is $P$ in this case).
Just as with the heat flux, $\bar{\sigma}$ determines the viscous heating rate and rate of momentum deposition, and its update is determined directly from its contribution to the momentum and energy equations.
Then, from the value of $\bar{\sigma}$, the local rate of change in $u(s)$ is given by the modified closure relationship
\begin{equation}
    \mu(T) \partial_s u = - \bar{\sigma} P / \sqrt{P^2 - \bar{\sigma}^2}.
\end{equation}
Similar to the heat flux closure, this prescription for the velocity gradient will be well-behaved so long as $\bar{\sigma}^2 < P^2$, which can be enforced during integration.

{Note that any of these alternate closures will necessarily change the asymptotic solution structure from the specific solution discussed above, and a new asymptotic solution (and corresponding boundary constraint) will have to be derived for each closure formulation.}


\section{Field-aligned Newtonian viscosity}\label{App::Vsc}

The form of the viscous stress used in Section \ref{Section::ODE} and its effect on the fluid are a useful simplification that has the appearance of an isotropic viscosity in 3D, while being approachable in 1D; however, that form cannot be derived from a field-aligned projection of either the isotropic or anisotropic limits of the 3D viscous stress as discussed by \citet{Spitzer:1953, Braginskii:1965, Holzer:1986a, Fitzpatrick:2015} and others. 
For the isotropic case there is no simple field-aligned projection unless we assume that the flow has no variation perpendicular to the magnetic field and that the field itself is uniform, which cannot be made consistent with the notion of a volume-filling solar wind.
For the anisotropic case, however, it is possible to derive the proper field-aligned projection by first beginning with the full 3D viscous stress and then constraining the direction of the flow.

The three viscosity-related source terms of interest are the total energy deposition rate,
\begin{equation}
    \dot{\mathcal{E}}_\mu = -\grad \cdot \left ( \uuline{\pi} \cdot {\mathbf u} \right),
\end{equation}
and the two separate contributions to this energy deposition rate: the viscous heating,
\begin{equation}
    \dot{\mathcal{Q}}_\mu = -\grad {\mathbf u} : \uuline{\pi},
\end{equation}
and the viscous force,
\begin{equation}
    \dot{\mbf{\mathcal{P}}}_\mu = - \grad \cdot \uuline{\pi}.
\end{equation}
Note, that the power (rate of work) associated with the viscous force is 
\begin{equation}
    {\mathbf u} \cdot \dot{\mbf{\mathcal{P}}}_\mu = - \mathbf{u} \cdot \left ( \grad \cdot \uuline{\pi} \right ) = \dot{\mathcal{E}_\mu} - \dot{\mathcal{Q}}_\mu,
\end{equation}
as expected. 
These expressions are independent of any particular closure; however, for this discussion we are interested in the classical (Newtonian) closure for an anisotropic magnetized plasma.

The general form of the anisotropic Newtonian viscous stress is given by
\begin{equation}
    \uuline{\pi} = -\uuline{\mu}^{(4)} : \uuline{W},
\end{equation}
where 
\begin{equation}
    \uuline{W} = \grad {\bf u} + (\grad {\bf u})^{T} - \frac{2}{3} \left ( \grad \cdot {\bf u} \right ) {\mathbb I}
\end{equation}
is the (second rank) rate of strain tensor, and 
\begin{equation}
    \uuline{\mu}^{(4)} = \mu \left ( \frac{3}{2} \bhat \bhat - \frac{1}{2} {\mathbb I} \right ) \bhat \bhat
\end{equation}
is the (fourth rank) viscosity tensor.
If we contract $\bhat\bhat$ with $\uuline{W}$ we can construct the projected rate of strain
\begin{align}
    W^{bb}({\bf u}) & = \hat{\mathbf{b}}\hat{\mathbf{b}}:
    \left ( \grad {\bf u} + (\grad {\bf u})^{T} - \frac{2}{3} {\mathbb I} \grad \cdot {\bf u} \right ) \notag \\
    & = 2 \left (  \bhat \cdot \left ( \bhat \cdot \grad \right ){\mathbf u} - \frac{1}{3} \grad \cdot {\mathbf u} \right ),
\end{align}
in terms of which the viscous stress is
\begin{equation}
    \uuline{\pi} = - \left ( \frac{3}{2} \bhat\bhat - \frac{1}{2}\mathbb{I}\right) \mu W^{bb} ({\bf u}).
\end{equation}
Here, the coefficient of kinematic viscosity, $\mu$, is understood to be a function of temperature, but further specification is not necessary for the present discussion.

As we are interested in the manifestation of viscosity in the context of field-aligned flows, we assume that the flow is along the unit vector of the magnetic field $\bhat$, and we write the velocity vector as ${\mathbf u} \rightarrow u \bhat$. 
We can then define the auxiliary variable $\eta$, given by
\begin{align} \label{Eqn::eta_aniso}
    \eta & = -\mu W^{bb}({\bf u} \rightarrow u \bhat) \notag \\
    & = - 2\mu \left (\bhat \cdot (\bhat \cdot \grad) u \bhat) - \frac{1}{3} \grad \cdot u \bhat \right) \notag \\
    & = - \frac{4}{3}\mu \left ( \partial_s u - \frac{1}{2} u \partial_s \ln A \right ),
\end{align}
where $\partial_s := \bhat \cdot \grad$ is the directional derivative along $\bhat$ and $\partial_s \ln A = \grad \cdot \bhat$, with $A \propto 1/|\mathbf B|$.

In terms of $\eta$ the parallel-flow limit of the viscous stress becomes
\begin{equation}
    \uuline{\pi}^{bb} = \left (\frac{3}{2} \bhat \bhat - \frac{1}{2} \mathbb{I} \right ) \eta,
\end{equation}
from which we can calculate the various contributions to the viscous heating, work, and total energy input.
For example, the inner product of the velocity with the viscous stress is given by
\begin{align}
    u \bhat \cdot \uuline{\pi}^{bb} & = u \bhat \cdot \left ( \frac{3}{2} \bhat\bhat - \frac{1}{2}\mathbb{I}\right ) \eta \notag \\
    & = \bhat u \eta,
\end{align}
from which it follows that the total viscous energy deposition rate is
\begin{align}
    \dot{\mathcal{E}}_\mu(u\bhat) & = - \grad \cdot \left ( \bhat u \eta \right ) \notag \\
    & = -\bhat \cdot \grad \left ( u \eta \right ) - u \eta \grad \cdot \bhat \notag \\
    & = -\partial_s (u \eta) - u \eta \partial_s \ln A \notag \\
    & = - {\rm div}(u \eta),
\end{align}
where ${\rm div} (u \eta) := (1/A) \partial_s (u A \eta) = \partial_s u \eta + u \eta \partial_s \ln A$.

We can similarly work through the viscous heating rate, which is given by:
\begin{align}
    \dot{\mathcal{Q}}_\mu(u\bhat) & = - \grad {\mathbf u}_{||} : \uuline{\pi}^{bb} \notag \\
    & = -\frac{\eta}{2} \left ( 3 \bhat \bhat - {\mathbf I} \right ) : \left ( \bhat \grad u + u \grad \bhat \right ) \notag \\
    & = -\frac{\eta}{2} \left (
    3 \bhat \cdot \grad u + 3 u \bhat \cdot (\bhat \cdot \grad) \bhat - \bhat \cdot \grad u - u \grad \cdot \bhat \right ) \notag \\
    & = -\frac{\eta}{2} \left ( 3 \partial_s u - \partial_s u - u \partial_s \ln A \right ) \notag \\ 
    & = -\eta \left ( \partial_s u - \frac{1}{2} u \partial_s \ln A \right ).
\end{align}

Finally, we can evaluate the component of the viscous force (density) in the direction of $\bhat$, which is given by
\begin{align}
    \dot{{\mathcal{P}}}_{\mu}(u\bhat) & = - \bhat \cdot \left ( \grad \cdot \uuline{\pi}^{bb} \right ) \notag\\
    & = - \grad \cdot \left ( \bhat \cdot \uuline{\pi}^{bb} \right ) + \uuline{\pi}^{bb}:\grad \bhat \notag \\
    & = - \grad \cdot \left ( \bhat \eta \right ) + \eta \left ( \frac{3}{2} \bhat \cdot ( \bhat \cdot \grad ) \bhat - \frac{1}{2}\grad \cdot \bhat \right ) \notag\\
    & = - \left( {\rm div} (\eta) + \frac{1}{2} \eta \partial_s \ln A \right ).
\end{align}

These equations for field-parallel flow limits of $\dot{E}_\mu$, $\dot{Q}_\mu$, and $\dot{\mbf{\mathcal{P}}}_\mu$, along with the definition of $\eta$ from equation \eqref{Eqn::eta_aniso} and specification of $\mu(T)$, fully specify the effects of classic (Newtonian) anisotropic viscosity for the case of plasma flow parallel to the magnetic field.

{Note that, as with the alternate closure relationships discussed above, these formulations of the viscous stress and corresponding force density will alter the asymptotic solutions, which will need to be re-derived to account for the various additional terms. See, e.g., the asymptotic solutions discussed by \citet{Summers:1980} and references therein.}


\end{document}